\documentclass[12pt,preprint]{aastex}

\usepackage{multirow}
\usepackage{natbib}
\usepackage{hyperref}
\usepackage[stable]{footmisc}
\newcommand{\myemail}{j.bagdonaite@vu.nl}

\shorttitle{Analysis of Molecular Hydrogen Absorption toward B0642$-$5038 for a Varying Proton-to-Electron Mass Ratio}
\shortauthors{Bagdonaite et al.}

\begin{document}


\renewcommand\thefootnote{\fnsymbol{footnote}}

\title{Analysis of Molecular Hydrogen Absorption toward \\ QSO B0642$-$5038 for a Varying Proton-to-Electron Mass Ratio$^*$}
\footnotetext[1]{Based on data obtained with UVES at the Very Large Telescope of the European Southern Observatory.}


\author{J. Bagdonaite$^{\dagger}$, W. Ubachs}\footnotetext[2]{{\myemail}}
\affil{Department of Physics and Astronomy, VU University Amsterdam, De Boelelaan 1081, 1081 HV Amsterdam, The Netherlands}
\and
\author{M.~T. Murphy, J.~B. Whitmore}
\affil{Centre for Astrophysics and Computing, Swinburne University of Technology, Victoria, 3122, Australia}




\renewcommand\thefootnote{\arabic{footnote}}

\begin{abstract}
Rovibronic molecular hydrogen (H$_2$) transitions at redshift $z_{\rm abs} \simeq 2.659$ towards the background quasar B0642$-$5038 are examined for a possible cosmological variation in the proton-to-electron mass ratio, $\mu$. We utilise an archival spectrum from the Very Large Telescope/Ultraviolet and Visual Echelle Spectrograph with a signal-to-noise ratio of $\sim$35 per 2.5-km$\,$s$^{-1}$ pixel at the observed H$_2$ wavelengths (335--410 nm). Some 111 H$_2$ transitions in the Lyman and Werner bands have been identified in the damped Lyman $\alpha$ system for which a kinetic gas temperature of $\sim$84 K and a molecular fraction $\log f = -2.18\pm0.08$ is determined. The H$_2$ absorption lines are included in a comprehensive fitting method, which allows us to extract a constraint on a variation of the proton-electron mass ratio, $\Delta\mu/\mu$, from all transitions at once. We obtain $\Delta\mu/\mu = (17.1 \pm 4.5_{\rm stat} \pm3.7_{\rm sys})\times10^{-6}$. However, we find evidence that this 
measurement has been affected by wavelength miscalibration errors recently identified in UVES. A correction based on observations of objects with solar-like spectra gives a smaller $\Delta\mu/\mu$ value and contributes to a larger systematic uncertainty: $\Delta\mu/\mu = (12.7 \pm 4.5_{\rm stat} \pm4.2_{\rm sys})\times10^{-6}$.

\end{abstract}





\section{Introduction}

The fact that only 4 $\%$ of the energy density content in the universe can be explained within the current framework of particle physics suggests that the Standard Model is incomplete. In this context, observational attempts to detect variation of fundamental constants serve as one of the guiding tools for theoretical extensions in both cosmology and the Standard Model. For instance, the theories which intend to unify gravitation with the other three fundamental forces by introducing additional spatial dimensions accommodate cosmologically varying constants quite naturally \citep{uzan2011}. Alternatively, additional quantum fields such as the dilaton field may be invoked predicting scenarios of varying constants complying with conservation of energy \citep{bekenstein2002}. Experimental methods are established for probing dimensionless constants such as the fine structure constant $\alpha\equiv e^2/(4\pi\epsilon_0 \hbar c)$, which sets the quantum electrodynamical scale, and the proton-to-electron mass ratio 
$\mu \equiv m_{\rm p}/m_{\rm e}$, which is sensitive to the ratio of the chromodynamic scale to the electroweak scale \citep{flambaum2004}.

The redshifted spectra of quasars contain multiple absorption lines arising from cold gas along the line of sight to Earth. These absorption lines serve as a means to probe the high redshift Universe, also making it possible to study varying fundamental constants. In the case of varying $\mu$ and $\alpha$, the absorbers which have the highest neutral hydrogen column densities, known as the Damped Lyman-$\alpha$ (DLA) systems, are of particular interest since they are most likely to contain molecular and atomic species which have $\mu$- or $\alpha$-sensitive energy levels. If $\mu$ or $\alpha$ changes with redshift, the relative pattern of the transitions, which is known to a very high accuracy from laboratory experiments, is altered in a specific and calculable manner, different from the overall redshift. 

A variation of $\mu$ can be detected through the Lyman and Werner bands of molecular hydrogen (H$_2$, $\lambda_{\rm rest}<1150$ \textrm{\AA}), that are observable from ground-based observatories if absorption occurs at redshifts $z_{\rm abs}>2$ \citep{thompson1975}. The proton-to-electron mass ratio at high redshift, $\mu_z$, is measured as a shift with respect to the present day value $\mu_0$: $\Delta\mu/\mu = (\mu_z-\mu_{0})/\mu_0$. H$_2$ absorption is detected quite rarely -- so far in some twenty DLAs, only few of which are suitable for high accuracy $\mu$ analysis, i.e. providing $\Delta\mu/\mu$ constraints at the level of  $< 10^{-5}$ \citep{ubachs2011}. For comparison, the atomic species (Mg, Fe, Cr, Zn etc) sensitive to a variation of $\alpha$ have been detected in some 300 absorption systems which has allowed the constant to be mapped over a broad spatial and temporal range and yielded an indication of spatial variation \citep{king2012}. As for $\mu$, the H$_2$ absorbing systems toward Q0347$-$383 
at $z_{\rm 
abs}=3.02$, Q0405$-$443 at $z_{\rm abs}=2.59$, Q0528$-$250 at $z_{\rm abs}=2.81$, J2123$-$005 at $z_{\rm abs}=2.06$, Q2348$-$011 at $z_{\rm abs}=2.42$, and HE0027$-$1836 at $z_{\rm abs}=2.40$ have been analysed thus far at high accuracy \citep{king2008, king2011, malec2010a, weerdenburg2012, wendt2012, bagdonaite2012, rahmani2013}. As a result of improving data quality and discussions on appropriate analysis methods, most of the absorbers have been studied more than once, sometimes leading to controversial outcomes. E.g the study by \cite{ivanchik2005} on Q0347$-$383 and Q0405$-$443 yielded indication of $\mu$ variation based on a laboratory based H$_2$ absorption spectrum using classical spectroscopy. An updated extreme ultraviolet laser spectrum of H$_2$ by \cite{reinhold2006} provided extended evidence for such a possible variation of $\mu$. However, a re-analysis of the astrophysical spectra of Q0347$-$383 and Q0405$-$443 by \cite{king2008}, using the so-called comprehensive fitting analysis also 
adopted in the present study, lowered the significance of the effect to a $<2\sigma$ deviation: $\Delta\mu/\mu=$(8.2$\pm$7.4)$\times 10^{-6}$ and (10.1$\pm$6.2)$\times 10^{-6}$ in the two sightlines, respectively.

In contrast, observations of the J2123$-$005 with Keck/HIRES delivered a $\mu$ constraint \citep{malec2010a} which was later reproduced by an independent revision (but same spectral analysis method) using a VLT/UVES spectrum of the same object \citep{weerdenburg2012}. Altogether, the H$_2$ studies converge in a constraint on the variation of $\mu$ at the level of $\Delta\mu/\mu< 1\times10^{-5}$ in the redshift range $z =2-3$. Given the sensitivity coefficients of H$_2$ ranging from $-0.02$ to $+0.05$, a substantially higher accuracy in $\Delta\mu/\mu$ can only be achieved by drastically improving the quality (signal-to-noise) of the spectra or by increasing the number of absorbers. The same holds true for the other two weak shifters: HD and CO molecules \citep{ivanov2008, salumbides2012}. 

As an alternative to H$_2$, inversion transitions of ammonia and rotational transitions of methanol can be used for they are, respectively, 2 and 3 orders of magnitude more sensitive to $\Delta\mu/\mu$ variation if compared to H$_2$ \citep{flambaum2007, jansen2011a, levshakov2011}. Detections of ammonia at redshifts $z_{\rm abs}= 0.89$ and 0.69 are known presently, yielding 1$\sigma$ constraints on $\mu$ at the level of $(1.0\pm4.7)\times10^{-7}$ and $(-3.5\pm1.2)\times10^{-7}$ \citep{henkel2009, kanekar2011}. Further observations of the object at redshift $z_{\rm abs} = 0.89$ -- a lensing galaxy toward the quasar PKS1830$-$211 -- has yielded a detection of methanol resulting in a stringent constraint of $\Delta\mu/\mu = (0.0\pm1.0)\times10^{-7}$ \citep{bagdonaite2013a}. Although ammonia and methanol are more favorable probes because of high sensitivity, their detections are extremely rare. Thus, molecular hydrogen remains a target molecule for $\mu$ variation studies, especially at high redshifts where no 
constraints from ammonia or methanol are 
available yet.  

In the study presented here, molecular hydrogen transitions at redshift $z_{\rm abs} = 2.659$ towards the background quasar B0642$-$5038 are analysed in search for a cosmological variation in the proton-to-electron mass ratio, $\mu$.

\section{Method}

\subsection{Molecular hydrogen}
For the $i$-th transition of H$_2$ detected in a cloud at redshift $z_{\rm abs}$, the observed wavelength is expressed as: 
\begin{equation}
\lambda_i = \lambda^{0}_{i}(1+z_{\rm abs})(1+K_i\frac{\Delta\mu}{\mu}),  
\label{eq1}                                                                                                                                                                                                                                                                                                                                                                                                                                                                                                                                                                                                                                                                                                          \end{equation}
where $\lambda^{0}_{i}$ is the rest wavelength of the transition, and $K_i$ is a sensitivity coefficient of the transition, expressing its shifting power and direction due to varying $\mu$. As for H$_2$ transitions, we employ sensitivity coefficients $K_i$ defined as 
\begin{equation}
K_i =\frac{\mu}{\lambda_i}\frac{d\lambda_i}{d\mu}\,.
\end{equation}
Note that this definition yields a different sign than usually adopted for the transitions in radio domain, where a relation is defined in terms of frequency: $\Delta\nu/\nu = K_i\Delta\mu/\mu$.
Nevertheless the sign of the $\Delta\mu/\mu$ remains unaffected, i.e. direct comparison of the optical and radio constraints is possible. From Eq. (\ref{eq1}) it follows that from two transitions of different sensitivities, one can determine both the redshift and $\Delta\mu/\mu$. In practice, it is desirable to use as many H$_2$ transitions as possible since the cumulative signal from multiple transitions is necessary to balance relatively small sensitivity coefficients. Compared to the uncertainties of line positions in the quasar spectrum, the laboratory wavelengths of H$_2$ at accuracies $\Delta\lambda/\lambda \sim 5\times 10^{-8}$ can be considered exact for our purpose \citep{salumbides2011}. The $K_i$ coefficients of H$_2$ have been calculated within a semi-empirical framework (\cite{ubachs2007}, used in present analysis), including effects beyond the Born-Oppenheimer approximation, and from ab initio methods \citep{meshkov2006}. The $K_i$ values from these two approaches are in agreement within 
1 $\%$. A comprehensive list of laboratory wavelengths, $K_i$ coefficients, oscillator strengths and damping parameters of H$_2$/HD transitions is provided by \cite{malec2010b} and implemented in this work (also see Fig. \ref{fig1}).

\subsection{Fitting method}
\label{method}
In this study we employ a fitting technique which relies on a number of physical assumptions allowing for simultaneous modelling of all H$_2$ transitions and nearby lying H\,\textsc{i} lines at once. This method is known as the ``comprehensive fitting method'', to distinguish it from a ``line-by-line'' method, in which, as its name suggests, each transition is fitted independently from the others. The former technique allows us to include more transitions than the latter, as the fitting of the surrounding Lyman-$\alpha$ forest is readily possible. Also, the line-by-line method may not be applicable for the H$_2$ absorbers with complex profiles (multiple H$_2$ clouds distributed close to each other in the velocity space). A more detailed comparison of these two fitting approaches is outlined by \cite{king2008} and \cite{malec2010a}.

In short, the analysis of H$_2$ absorption spectra can be described as a three-step process:
\begin{enumerate}
 \item Selection of potentially usable transitions. 
 \item Setting up and refining of the fit. 
 \item $\mu$ derivation and testing.
 
\end{enumerate}

The analysis starts from visual inspection of the spectrum. The goal is to compose a list of spectral segments each containing an  H$_2$ line and some surrounding buffer region so that any overlapping non-H$_2$ feature can be modeled too. Quite frequently two or more adjacent H$_2$ lines are included in a single region when they are too close to be fitted separately. Being distributed over the Lyman-$\alpha$ forest, most of the H$_2$ transitions unavoidably overlap with H\,\textsc{i} lines. In some cases a strongly saturated H\,\textsc{i} absorption may render overlapping H$_2$ line(s) useless, in which case they are neglected. Additionally, some H$_2$ transitions can be excluded because of an overlap with narrow metal or unidentified lines if these are unconstrainable (see Section \ref{section:abs-model}). The number of selected H$_2$ lines is usually in the range between 40 and 100.

Once the list is complete, the non-linear least squares Voigt profile fitting program \textsc{vpfit} 9.5\footnote{Developed by B. Carswell et al.; \url{http://www.ast.cam.ac.uk/~rfc/vpfit.html}. We use an upgraded version of \textsc{vpfit} 9.5 in which the Gauss-Newton optimisation algorithm is augmented with the Levenberg-Marquardt algorithm, and computing can be done in parallel on multiple cores. These changes were implemented by J. King (UNSW).} is used to model the absorption lines. A Voigt profile represents an absorption line which is broadened by Doppler and Lorentzian mechanisms, and convolved with an instrumental profile (assumed to be Gaussian). The Doppler broadening is caused by the thermal (or large-scale turbulent) motion of the molecules/atoms, while the Lorentzian broadening is due to the finite lifetimes of the excited states. All H$_2$ and surrounding neutral hydrogen lines are modelled simultaneously. For each transition, a Voigt profile is assigned by providing three adjustable 
parameters: the redshift 
of the transition, $z$, the column density, $N$, and the velocity width, $b$. Initial user-supplied parameter guesses for the considered lines are fed to \textsc{vpfit}, which finds the best match between the model and data by minimizing the goodness-of-fit parameter $\chi^2$. At each iteration, \textsc{vpfit} checks the change in the relative $\chi^2$ and reports convergence once an improvement tolerance is met. The stopping criterion, the stepping size for each of the free parameters and their limits are user-defined. Further, the model is refined manually by adding or removing lines (i.e. H$_2$, H{\sc \,i} Lyman-$\alpha$ and/or metallic ion transitions), and again fitted with \textsc{vpfit}. This process is repeated until a statistically acceptable fit is achieved. There are several main guidelines for a statistically acceptable model:
 \begin{itemize}
 \item The residuals of each fitted region should be well behaved, that is they should be free from nonrandom trends. This is verified by inspecting each individual region by eye and by constructing a composite residual spectrum (CRS) which is an even more sensitive tool to systematic fitting problems. To compose a CRS, we select the cleanest H$_2$ transitions, normalize their residuals (fit minus data) by the flux error, and shift them to a common redshift/velocity scale where they are averaged together. If any systematic underfitting is present, it becomes `amplified' in the CRS.
 \item A statistically adequate fit should have a $\chi^2$ per degree of freedom, $\chi^2_{\nu}$, around 1. For the entire fit with $\nu$ degrees of freedom\footnote{The degrees of freedom in \textsc{vpfit} are assumed to be $\nu = n-p$ where $n$ is the number of data points fitted and $p$ is the number of free parameters. However, $\nu$ is not defined in such way for a non-linear model. The rationale for this assumption was discussed by \cite{king2012thesis}.}, a $\chi^2/\nu$ is reported by \textsc{vpfit}. Competing models can be rated according to their relative values of $\chi^2_{\nu}$. 
 \item The statistically most adequate model has the lowest Akaike Information Criterion (for finite sample sizes abbreviated AICC, \cite{akaike1974}). The AICC is a method to approve or disprove the addition of absorption lines (or more often in our case -- velocity components of H$_2$). It is defined as: 
\begin{equation}
 {\rm AICC}=\chi^2 + 2p + \frac{2p(p+1)}{n-p-1}    
\end{equation} 
 where $n$ is the number of data points fitted and $p$ is the number of free parameters. When comparing two models, the model with a lower AICC is statistically preferred. More precisely, $\Delta$AICC$> 5$ is considered strong evidence and $\Delta$AICC$> 10$ is considered very strong evidence for choosing the model with a lower AICC.
\end{itemize}
 
The number of so called velocity components (VC) is an important ingredient in deciding which is the best model to represent a H$_2$ spectrum. The light coming from a quasar can be absorbed by multiple H$_2$ clouds in the DLA (or a single cloud can contain clumps with varying density, turbulence and temperature), resulting in multiple spectral features for every H$_2$ transition. At first, the number of velocity components is estimated by eye. Then, once the initial model is fitted with \textsc{vpfit}, more H$_2$ components might be added if the residuals show any hints of underfitting. \textsc{vpfit} rejects the added components if data do not support them. As explained above, the necessity to include more components is also assessed from the behaviour of $\chi^2_{\nu}$ and AICC, and the residuals -- if these parameters improve, the model is considered more adequate. One should be cautious to not to underfit: it has been shown that underfitting is more prone to cause biases in the $\Delta\mu/\mu$ or 
$\Delta\alpha/\alpha$ constraints than overfitting \citep{murphy2008a, murphy2008b}.

As mentioned before, each Voigt profile is described by 3 free parameters. By employing the comprehensive fitting method, we tie the free parameters among different transitions. For a single velocity component of H$_2$ the following physical restrictions are imposed: 
\begin{itemize}                                                                                                                                                                                                                                                                                                                                             \item same column density for all transitions from a single $J$ level 
\item same $z$ parameter for all transitions
\item same turbulent $b$ parameter for all transitions                                                                                                                                                                                                                                                                                                                               \end{itemize}
Some assumptions can be released if a reasonable fit cannot be achieved but, as a rule, we aim to base our fit on as much molecular physics information on the H$_2$ molecule as possible. Besides the free parameters, a Voigt profile for each transition involves 3 fixed parameters which are obtained from the molecular physics: rest wavelength $\lambda_i$, oscillator strength $I_i$, and damping parameter $\Gamma_i$.

Once the model is optimized, the last free parameter -- $\Delta\mu/\mu$ -- is introduced. A single $\Delta\mu/\mu$ parameter allows relative shifting of the H$_2$ transitions in accordance with Eq. (\ref{eq1}). It is important to introduce this fourth parameter only after the model is finalized because otherwise it can acquire a false value to accommodate imperfections of the fit. The very last step is to test how sensitive the derived $\Delta\mu/\mu$ constraint is to the physical assumptions and fitting choices made. These tests can include $\mu$ constraints derived from various data cuts: separate $J$ levels, the Werner transitions and Lyman transitions separately, regions with no overlapping metal lines, a fit with no assumption about $b$ in different $J$ levels, etc.

\section{Data}

The spectrum of the QSO B0642$-$5038 (RA $06^{h}43^{m}27^{s}.0024$ Dec -$50^{\circ}41\arcmin12\arcsec$.804, J2000, visible magnitude V=18.5) was obtained on VLT/UVES under three different programmes (their IDs along with the observational settings are listed in Table \ref{observations}). A concise report on the observations was given by \cite{noterdaeme2008}. All raw science and calibration data were retrieved from the publicly available ESO archive\footnote{ESO Archive is accessible via: \url{http://archive.eso.org/eso/eso_archive_main.html}}. The total exposure time of the selected data makes up 22.4 hrs. The combined spectrum covers wavelengths from 330 to 1040 nm with two gaps due to separation between the CCDs at 452.1--461.9 nm and 842.5--856.9\,nm. The ratio of average seeing to slitwidth is 1.1. More than half of the exposures were followed immediately by the ThAr calibration (so-called `attached' ThAr calibration) thereby minimizing the possibility that the optical system can be disturbed by the 
grating reset. Most of the science frames were obtained with a slitwidth of 1.0$\arcsec$ which, with 2$\times$2 on-chip binning, translates to an instrumental velocity resolution of $\sigma_{v} \simeq 3.1$ km s$^{-1}$ in the blue part and $\simeq$3.3 km s$^{-1}$ in the red part (resolving powers R$\sim$41 000 and 39 000, respectively). However, the target is a point source and, unlike the calibration lamp which illuminates the entire slit, it only covers the central part of the slit, thus a better resolution is expected. Here, we adopt a resolution of $\sigma_{v} \simeq 2.9$ km s$^{-1}$ which is $\sim$6$\%$ better than the formally derived value.

We used the ESO UVES Common Pipeline Language software suite to optimally extract and calibrate the echelle orders, and an open-source, publicly available, custom program \textsc{uves$\_$popler}\footnote{\url{http://astronomy.swin.edu.au/~mmurphy/UVES_popler}}
 to combine multiple exposures into a single, 1D, normalised spectrum on a vacuum-heliocentric wavelength scale. The Pipeline first bias-corrects and flat-fields the quasar exposures. It then constrains a physical model of where the quasar light is expected to fall on the CCDs with the aid of quartz and ThAr exposures taken through a "pinhole" instead of a standard slit. 
 The quasar light is then separated from the background and extracted using an optimal extraction method. The ThAr flux (from the calibration frame taken with the same slit as the quasar) was extracted with the same optimal weights derived from the corresponding quasar exposure. The selection of ThAr lines and the establishment of a wavelength solution followed the method described in \cite{murphy2007}. \textsc{uves$\_$popler} takes the pixel-space flux spectra from each order, of each exposure, and disperses all onto a common log-linear wavelength scale with a dispersion bin of 2.5 km s$^{-1}$. 
 Before redispersion, the wavelength solution for each exposure (established from its corresponding ThAr calibration frame) was corrected from air to vacuum using the \cite{edlen1966} formula and placed in the heliocentric reference frame using the time of the mid-point of the exposure’s integration as a reference. 
 The method of combining the exposures is based on the relative flux scaling between all available overlapping orders weighted by their inverse-variance. When combining these scaled orders in this weighthed fashion, \textsc{uves$\_$popler} applies an automated algorithm for cosmic ray rejection while other spectral artefacts are inspected and removed manually. The continuum is fitted manually with low-order polynomials. Small constant or linear local changes of the continuum and corrections to zero level are permitted in later stages of the analysis.

\begin{table}[h!]

\caption{\small{ESO archival data of the B0642$-$5038 and Ceres observations with
VLT/UVES. The total combined integration time on the B0642$-$5038 makes up 22.4 hrs in each arm,
14.4 hrs of which have the attached ThAr calibration. The CCD binning mode was
2$\times$2 for all B0642$-$5038 frames. Airmasses were in the range between 1.113 and 1.562. For Ceres, HD28099, and HD76151 the binning mode was 1$\times$1.}}
\begin{tabular}{c c c c c c c}
\hline
Program ID & Obs. date & Blue/red & Slit& Integration & Seeing & ThAr \\
	    &	    & [nm]             & [arcsec] & time [s]   &  [arcsec]& calibration \\ \hline \hline
B0642$-$5038 & & & & & & \\ \hline
073.A-0071(A) &2004-09-17   &390/580   &1.2 &5500  & 1.89& non-att.  \\        
	      &2004-09-18   &390/580   &1.2 &6000& 1.90 &non-att.  \\
	      &2004-09-19   &390/580   &0.8 &6000& 1.00 &non-att.  \\ \hline
074.A-0201(A) &2004-10-09   &390/850   &0.8 &5800& 0.60 &non-att.    \\
	      &2004-10-10   &390/850    &0.8 &5500& 0.37 &non-att.  \\ \hline   
080.A-0288(A)  &2007-12-11    &390/564   &1.0 &3725&1.00 &att.  \\	    
	      &2008-01-03     &390/564   &1.0 &3725 & 0.91&att.  \\
	      &2008-01-04     &390/564   &1.0 &3725&1.77 &att.  \\
	      &2008-01-04     &390/564   &1.0 &3725&1.66 &att.   \\
	      &2008-01-04     &390/564   &1.0 &1389&1.69 &att.  \\
	      &2008-01-06     &390/564   &1.0 &2104&1.24 &att.   \\
	      &2008-01-06     &390/564   &1.0 &3725&1.24 &att.  \\
	      &2008-01-06     &390/564   &1.0 &3725&1.08 &att.  \\
	      &2008-01-07     &390/564   &1.0 &3725&0.84 &att.  \\
	      &2008-01-13     &390/564   &1.0 &3725&1.00 &att.  \\
	      &2008-01-16     &390/564   &1.0 &3725&0.85 &att.  \\
	      &2008-02-06     &390/564   &1.0 &3725&0.82 &att.  \\
	      &2008-02-07     &390/564   &1.0 &3725&0.90 &att.  \\
	      &2008-02-09     &390/564   &1.0 &3725&0.80 &att.  \\
	      &2008-02-10     &390/564   &1.0 &3725&0.81 &att.  \\ \hline
CERES & & & & & & \\ \hline
080.C-0881(B) &2007-12-05 &  346 & 1.0 &2850 &1.32  &non-att.  \\ \hline
HD28099 & & & & & & \\ \hline
380.C-0773(A) &2008-01-11 & 390/564  	& 0.7 & 264 &1.27  & non-att. \\
	      &2008-01-11 & 390/564 	& 0.7 & 264 &1.21  & non-att. \\ \hline
HD76151 & & & & & & \\ \hline
380.C-0773(A) &2008-01-11 & 390/564	& 0.7 & 67 &0.69  &att.  \\ \hline
\end{tabular}
\label{observations}
\end{table}

\section{Analysis}

\subsection{Damped Lyman Alpha system in the QSO B0642$-$5038 spectrum}
The quasar is located at redshift $z=3.09$ thereby defining the extent of the Lyman-$\alpha$ forest. The quasar radiation is cut off at the shortest recorded wavelengths since a DLA at redshift $z=$2.659 produces the Lyman break at $<$335 nm. The H$_2$ lines associated with the DLA are spread in the range from 335 to 410 nm. The signal-to-noise ratio of the continuum at the center of this range ($\sim$370 nm) is 35 per 2.5-km$\,$s$^{-1}$ pixel. The column density of the neutral hydrogen contained by the DLA is $\log N$ = 20.95$\pm$0.08 cm$^{-2}$ \citep{noterdaeme2008}. 

\begin{figure}[h!]
\centering
 \includegraphics[scale=1.1]{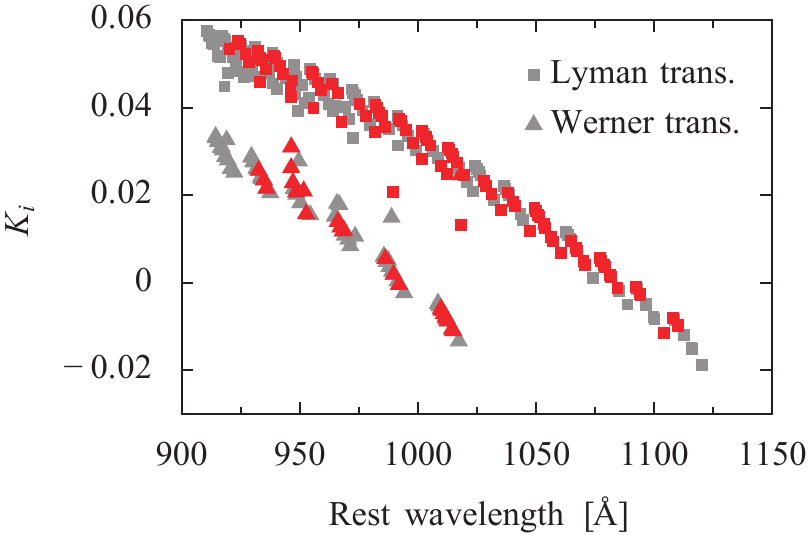}
\caption{Sensitivity coefficients $K_i$ of the Lyman and Werner transitions of H$_2$ plotted against the rest wavelength. The transitions which were suitable for the $\Delta\mu/\mu$ analysis are marked in red.}
\label{fig1}
\end{figure}

\begin{figure}[h!]
\centering
 \includegraphics[scale=1.0]{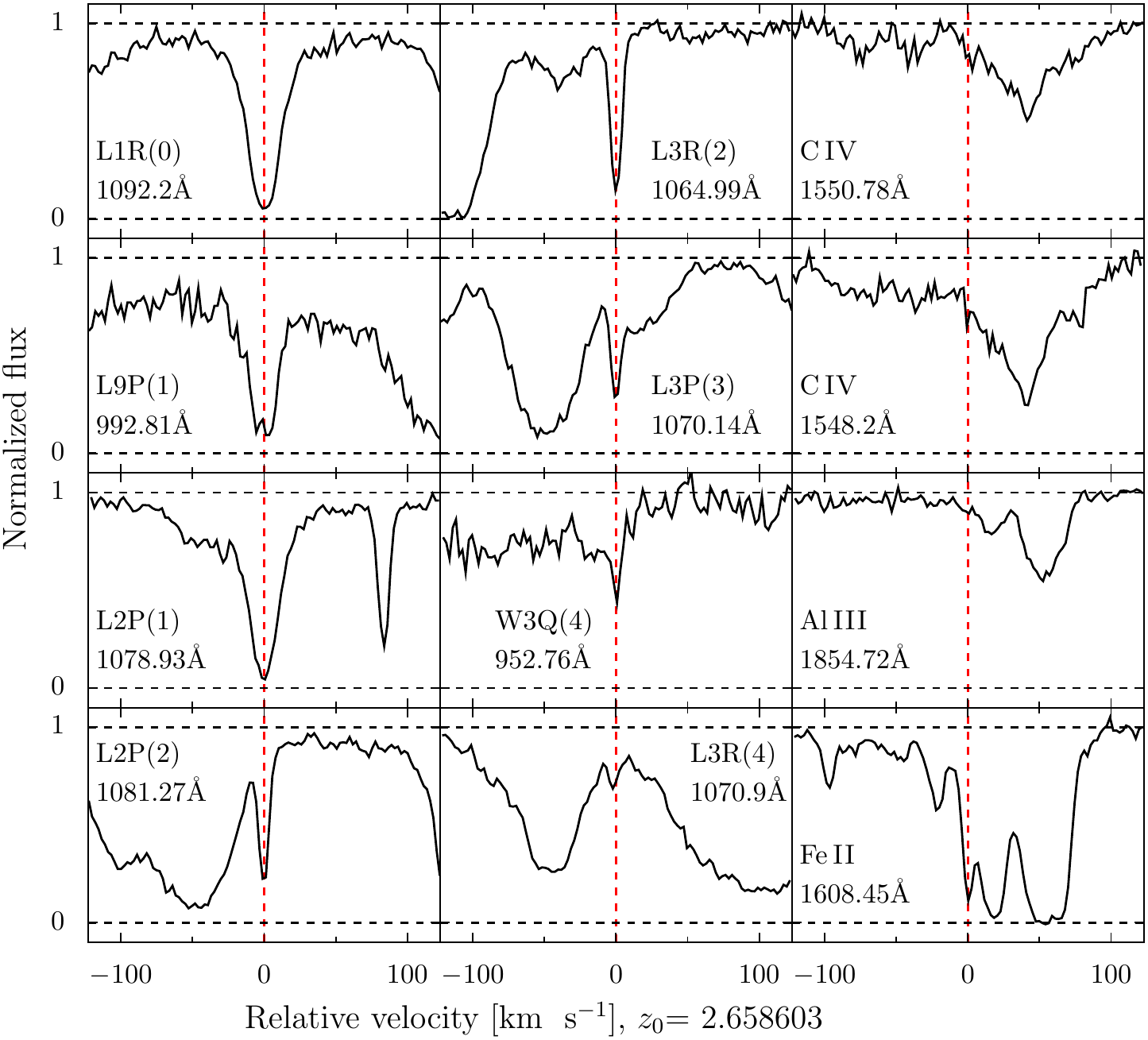}
\caption{Some of the H$_2$ and metal transitions associated with the DLA at $z \sim 2.659$, displayed on a velocity scale. The red dashed line is centered at the redshift of H$_2$ absorption. The C\,$\textsc{iv}$, Al\,$\textsc{iii}$ and Fe\,$\textsc{ii}$ profiles show multiple absorption features spread over 200 km s$^{-1}$.}
\label{fig0}
\end{figure}

\begin{figure}[h!]
\centering
 \includegraphics[scale=0.9]{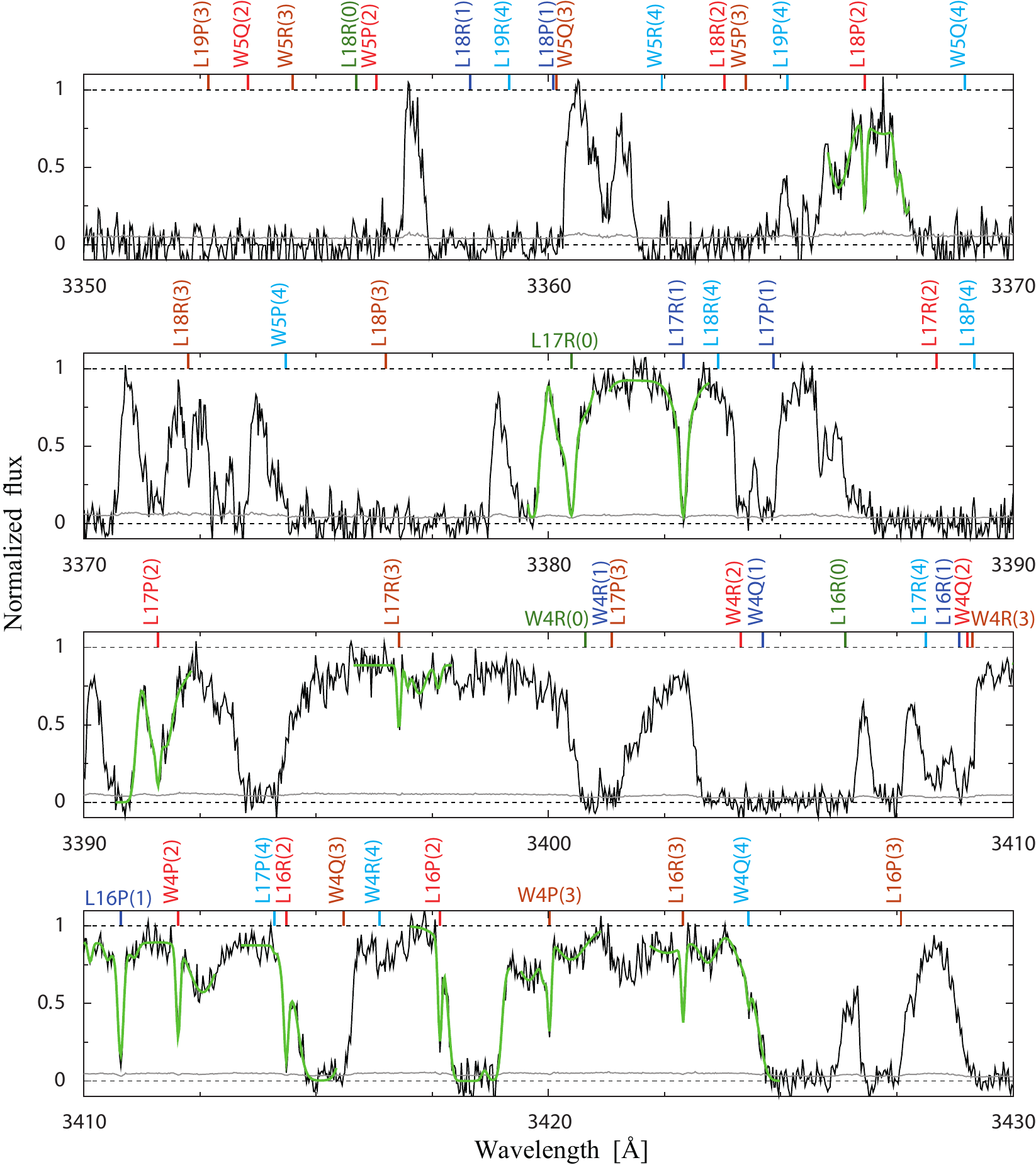}
\caption{Part of the B0642$-$5038 spectrum with fitted H$_2$ absorptions. The tickmarks are positioned at $z=2.658603$. Different $J$ transitions are labeled in different colors. The photons at the bluest wavelengths are cut out by the Lyman limit of the DLA.}
\label{fig3}
\end{figure}

\subsection{Creating and selecting the most adequate absorption model}
\label{section:abs-model}
We detect molecular hydrogen transitions up to rotational levels with $J=4$. In total there are $\sim$250 potentially useful Lyman and Werner transitions from $J\leq4$ and vibrational levels $v\leq18$ (Lyman) and $v\leq5$ (Werner). We select those H$_2$ transitions which are neither overlapped by saturated H$\,\textsc{i}$ nor too weak, as suitable for the analysis. Altogether, 72 regions have been selected, containing a total of 111 H$_2$ transitions (see Table \ref{list-transitions}; Fig. \ref{fig3} displays a part of the spectrum with selected transitions). Each of the H$_2$ regions is inspected for possible intervening metal lines. The redshifts of absorbing systems with metals in this sightline are determined by identifying the lines in the spectrum redwards from the Lyman-$\alpha$ emission of the quasar. Some C\,\textsc{iv} systems could be found bluewards too. All of them are listed in Table \ref{metals}. We identify seven H$_2$ transitions which are overlapped by metal lines, and several more which 
are situated next to some narrow unidentified lines. To constrain the relevant metal transitions three more regions containing their counterparts in the red part of the spectrum are added to the region sample. For instance, the H$_2$ transitions L2P(2) and L2R(3), redshifted to 3954-3958 \textrm{\AA}, are overlapping with a Fe\,$\textsc{ii}$ transition ($\lambda_{\rm rest}=1081.87$\,\textrm{\AA}) of the DLA. To constrain the absorption of Fe\,$\textsc{ii}$, we use a different Fe\,$\textsc{ii}$ transition which is outside the Lyman-$\alpha$ forest ($\lambda_{\rm rest}= 1608.45$\,\textrm{\AA}).

\begin{table}[h!]
\begin{minipage}{12 cm} 
\caption{A list of the H$_2$ transitions used in the present study. The underlined transitions are overlapped by narrow lines of known or unknown origin, and the corresponding regions are excluded in one of the tests.}  \smallskip
\begin{tabular}{c|l l| c} 

$J$ level & Lyman transitions   & Werner transitions & $n_{trans}$  \\ \hline

$J=0$ & L0R0, L1R0, \underline{L2R0}, L4R0, L7R0, & W3R0 & 9 \\
      & L8R0, L14R0, L17R0  & & \\
$J=1$ &  L0P1, L1P1, \underline{L2P1}, \underline{L2R1}, L4P1, & W0Q1, W2Q1, W3R1 & 25  \\
	 &  L4R1, L5P1, L7P1, L7R1, L8P1,     & & \\
	 &  L9P1, L9R1, L10P1, L10R1, \underline{L12R1},& & \\
         &  \underline{L13P1}, \underline{L13R1}, L14R1, L15P1, L15R1, & & \\
         &  L16P1, \underline{L17R1}, & & \\
$J=2$ &  L0R2, L1R2, \underline{L2P2}, \underline{L2R2}, L3P2,    &  W0Q2, W0P2, W1R2,  & 32 \\
	 &  L3R2, L4P2, L4R2, L5P2, L6P2,  & W2Q2, W2P2, W4P2, &\\ 
	 &  \underline{L7P2}, L7R2, L8P2, L8R2, L9P2,       &  W3R2 & \\
	 &  L10P2, \underline{L10R2}, \underline{L11P2}, L12P2, L13P2,     & & \\
         &  L15P2, L16P2, L16R2, L17P2, L18P2  & & \\
$J=3$ &  \underline{L2R3}, L2P3, L3P3, L3R3, L4P3,  & W0R3, W0Q3, W0P3,  & 25 \\
	 & L4R3, \underline{L5R3}, L6P3, L6R3, L7P3, & \underline{W1Q3}, W3R3, W3P3,   & \\ 
	 & L9P3, L10R3, L11P3, L12R3, L13R3, &W4P3 &\\
         & L15R3, L16R3, L17R3 & &\\
$J=4$ &  L1P4, L3R4, L4P4, L4R4, L5P4,   & W0Q4, W0R4, W1Q4, & 20  \\
      &  L6P4, L8P4, L8R4, L9P4, \underline{L10R4}, & W2R4, W3Q4, W4Q4   &\\
      &  L11P4, \underline{L14P4}, L15R4, L17P4  &  &\\
Total & & & 111    \\
\end{tabular}
\label{list-transitions}
\end{minipage}
\end{table}

Although the H$_2$ absorption is seen in a single feature upon first inspection (see Fig. \ref{fig0}), a possibility of a more complex underlying structure is explored (see Section \ref{method} for motivation of doing so). A second velocity component is added near to the 1st VC, and fitting is performed with \textsc{vpfit}. Several fits with various combinations of the initial values for $z$, $b$ and $N$ of the two components have been carried out to ensure robustness of such model. For each VC, the free parameters are connected to each other as described in Section \ref{method}. We find that the model with 2\,VC is statistically more preferable than the 1\,VC model: it has a $\chi^2_{\nu}$ of 1.189 and an AICC of 10165.6, compared to, respectively, 1.193 and 10190.4 of the 1\,VC model. However, the additional VC, present in $J=0$ and 1 transitions only, is very weak (a column density of some four orders of magnitude lower than the main component is found) and its position has a very large uncertainty (see 
Table \ref{list-columns}) which later translates to a minor increase in the uncertainty of $\Delta\mu/\mu$. The composite residual spectrum of the 1\,VC model does not show any significant underfitting (see Fig. \ref{fig2}). Attempts to compose a stable 3\,VC model were unsuccessful. Thus, based on the statistical parameters, we adopt the 2\,VC model as fiducial but we perform further testing for robustness and consistency on both 1\,VC and 2\,VC models, since the 2nd VC is very weak. 

\begin{figure}[!h]
 \includegraphics[scale=1.1]{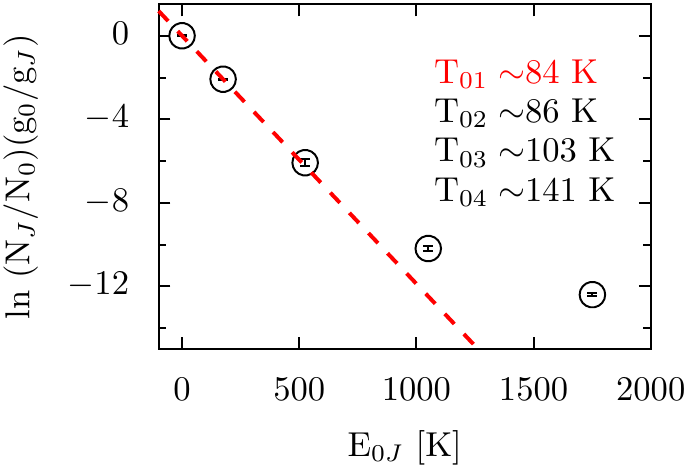}
 \caption{Excitation diagram for H$_2$ in the DLA at $z=2.659$. The column densities are weighted with a factor $g_J=g_N(2J+1)$, where $g_N$ is the nuclear spin weight that has value $g_N=1$ for even values of $J$ and $g_N=3$ for odd values of $J$. The slope of a straight line gives $1/T_{01}$. The uncertainties of $T_{0J}$, as derived directly from measured column densities, are of the order of 3-5 K.
}
\label{temperature}
\end{figure}

\begin{figure}[h!]
\centering
 \includegraphics[scale=1.2]{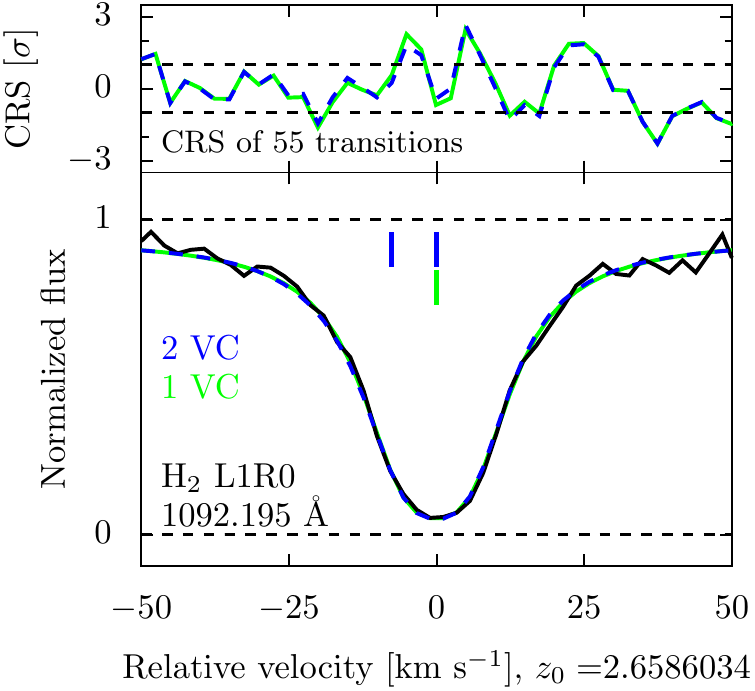}
\caption{The composite residual spectrum (top) and an example of H$_2$ absorption centered at $z=2.658603$. Fifty five H$_2$ transitions were used to compose the composite residual spectrum. The black horizontal dashed lines in the composite residual plot show $\pm1\sigma$ boundaries. The green line shows composite residuals (top) and a corresponding fit from the 1\,VC model while the blue line refers to the 2\,VC model.}
\label{fig2}
\end{figure}

Before proceeding with consistency tests focused on a $\mu$-variation analysis, we use the measured column densities of H$_2$ to estimate some basic characteristics of the absorbing cloud. The column densities of different $J$ transitions of H$_2$ provide a measure of the gas temperature in the cloud if the observed populations are in thermodynamic equilibrium and follow Boltzmann's law. The excitation temperature is defined with respect to the $J = 0$ level via: $N_J$/$N_0$ = $g_J/g_0\times e^{-E_{0J}/kT_{0J}}$. In Galactic diffuse clouds, the $T_{01}$ temperature is regarded as a kinetic temperature of the gas \citep{roy2006}. The higher levels are often populated in excess of the Boltzmann law, implying that besides collisions there are other processes involved such as cascades following UV or formation pumping or, alternatively, shocks or turbulence effects (see e.g. \cite{cechi2005}).
From the H$_2$ absorption in the DLA at $z=2.659$ toward B0642$-$5038, we find a kinetic temperature of $T_{01}\simeq84$ K, which is consistent with that in the Galactic ISM \citep{rachford2002}, while the levels at $J>2$ indeed show higher temperatures of 100-140 K (see Fig. \ref{temperature}). Both measured temperatures are consistent with what was found in other DLA studies (e.g. \cite{srianand2005}). Note that for the temperature calculations we used column densities from the 1\,VC model.

The total molecular column density amounts to $\log N$(H$_2$)= $18.48\pm0.01$ cm$^{-2}$. Given the neutral content of $\log N$(H\,\textsc{i})= $20.95\pm0.08$ cm$^{-2}$ \citep{noterdaeme2008}, a molecular fraction of $\log f = -2.18\pm0.08$ is derived. Only 15$\%$ of DLAs with detected H$_2$ have molecular fraction in the range $-4 < \log f < -1$ \citep{ledoux2003}. For comparison, in the Galactic sightlines of similar neutral hydrogen density, typical values for the molecular hydrogen are $\log N$(H$_2$) $>$ 19 \citep{rachford2002}.

\begin{table}[h!]
\caption{Metal absorbers in the line of sight toward the B0642$-$5038}
\begin{tabular}{c|c}
\hline
Redshift & Species \\ \hline         
 1.545\tablenotemark{a}, 1.647, 1.691, 1.852,   &  \\
 1.987, 2.031, 2.126, 2.204,  &  C\,\textsc{iv}\\
 2.348, 2.423, 2.500, 2.912  & \\
 1.561  & C\,\textsc{i}, C\,\textsc{iv} \\
 2.029 & C\,\textsc{iv}, Si\,\textsc{iv}, Si\,\textsc{iii} \\
 2.082 & C\,\textsc{iv}, Si\,\textsc{iv}, Si\,\textsc{iii} \\
 2.510 & C\,\textsc{iv}, N \textsc{v}, Si\,\textsc{iv}, O \textsc{vi} \\
 2.521 & C\,\textsc{iv}, Al\,\textsc{ii} \\
 2.659\tablenotemark{b} & C\,\textsc{iv}, C\,\textsc{ii}, Al\,\textsc{iii}, Si\,\textsc{ii}\tablenotemark{a}, Si\,\textsc{iii}, Si\,\textsc{iv}, P\,\textsc{ii}, \\
       & Cr\,\textsc{ii}, Ni\,\textsc{ii}, Zn\,\textsc{ii}, Al\,\textsc{ii}, Fe\,\textsc{ii}\tablenotemark{a}, N\,\textsc{i}, O\,\textsc{i}, C \textsc{I}, \\
       & C\,\textsc{iii}, C\,\textsc{ii}* \\
 2.899 & C\,\textsc{iv}\tablenotemark{a}, C\,\textsc{iii}\tablenotemark{a}, N \textsc{v}, Si\,\textsc{iv}, Si\,\textsc{iii}, S\,\textsc{iv}, O \textsc{vi} \\
 2.955 & C\,\textsc{iv}, C\,\textsc{iii}, O \textsc{vi} \\
 2.967 & C\,\textsc{iv}, O \textsc{vi} \\

\end{tabular}
\tablenotetext{a}{Metal absorptions overlapping H$_2$ (or those used to constrain them)}
\tablenotetext{b}{The DLA system presently analyzed for $\mu$-variation}
\label{metals}
\end{table}

\begin{table}[h!]
\begin{minipage}{12 cm} 
\caption{Column densities, Doppler widths and redshifts of the fitted H$_2$ transitions (together with 1$\sigma$ statistical uncertainties) as reported by \textsc{vpfit}.}  \smallskip
\begin{tabular}{|l|l|l|l|} \hline
\multirow{4}{*}{} & 1\,VC model & \multicolumn{2}{c|}{2\,VC model}  \\ \hline
&$z=2.6586026(4)$ & $z_1=2.6586030(4)$ &$z_2=2.65851(12)$\\
&$b=1.57\pm0.04$ km s$^{-1}$& $b_1=1.58\pm0.04$ km s$^{-1}$  & $b_2=12.5\pm6.3$ km s$^{-1}$  \\ \hline
$J$ level&$\log N$ [cm$^{-2}$] & $\log N$ [cm$^{-2}$] & $\log N$ [cm$^{-2}$] \\ \hline
$J=0$ & 18.15$\pm$0.01 &18.13$\pm$0.02 & 13.97$\pm$0.62 \\
$J=1$ & 18.20$\pm$0.01 & 18.19$\pm$0.01 & 13.74$\pm$0.61\\
$J=2$ & 16.21$\pm$0.07 & 16.19$\pm$0.07 &  -\\
$J=3$ & 15.04$\pm$0.05 & 15.03$\pm$0.05 & -\\
$J=4$ & 13.72$\pm$0.03 & 13.71$\pm$0.03 & -\\ \hline
\end{tabular}
\label{list-columns}
\end{minipage}
\end{table}

\subsection{Consistency tests}
\label{tests}
The results of the consistency tests discussed below are displayed in Fig. \ref{fig4}. We start from a generic fit that includes all suitable transitions (111 in total) and data, i.e. all exposures independent of how they were calibrated (attached ThAr or not). Unless stated differently, in all the tests the $z$ and $b$ parameters are assumed to be the same for all $J$ levels, and the column density $N$ is the same for all transitions from a single $J$ level for a single velocity component. The fitting solution is found through an iterative process, where at each iteration \textsc{vpfit} checks the change in the relative $\chi^2$ and stops when a stopping criterion is reached. We used a stopping criterion of $\Delta\chi^2/\chi^2  < 10^{-7}$ for a fine mode of fitting and $< 10^{-6}$ for a coarse fitting. Whenever possible, we used the fine fitting mode, but in some cases this gave very slow convergence of the fit and the coarse mode gave faster, more reliable results from a single minimization run of \textsc{
vpfit}. The 1\,VC generic model delivers $\Delta\mu/\mu$ = (16.9$\pm$4.4$_{\rm stat}$)$\times10^{-6}$ while the 2\,VC generic model results in $\Delta\mu/\mu$ = (17.1$\pm$4.5$_{\rm stat}$)$\times10^{-6}$; further tests and comparisons are made with respect to these results from generic tests.

In all the tests presented here, the transition oscillator strengths were kept fixed to the values established for the H$_2$ molecule. Note that the line intensities were derived in a model involving interactions between electronic states B$^1\Sigma_u^+$, C$^1\Pi_u$, B$'^1\Sigma_u^+$, and D$^1\Pi_u$ and intensity borrowing \citep{abgrall1994};
no accurate transition oscillator strengths have been experimentally determined. In contrast, in previous analyses by \cite{malec2010a} and \cite{king2011} for some specific
transitions the oscillator strengths had to be adapted in the fitting procedure. In a study of three H$_2$ absorbing systems \cite{king2008} varied column densities (including a product with an oscillator strength) for each of the lines separately. In the present study we stuck as close as possible to the established molecular physics of H$_2$ and found no need to adapt oscillator strengths.

\subsubsection{Isolating low- and high-$J$ transitions}
\label{low-high-J-test}
In this particular absorber, the $J=0,1$ and $J=2-4$ transitions differ in that the former are saturated ($\log N > 18$ cm$^{-2}$, Table \ref{list-columns}) unlike the latter. For this reason, if the velocity structure is underfitted (velocity components missing), the lower $J$ level transitions would be affected more than the higher $J$ transitions because any missing components will be weak even in the low $J$ transitions and completely negligible in the high $J$ transitions. In other words, the low $J$ transitions can be more misleading when determining $\Delta\mu/\mu$ than the high $J$ transitions. For this reason, two more $\Delta \mu/\mu$ constraints are derived separately from the $J = 0,1$ transitions (34 in total) and $J = 2-4$ transitions (77 in total) of the generic fit. Note, that the tests are performed not by excluding unwanted transitions but by leaving their $\Delta\mu/\mu$ parameter fixed to zero. As in the generic tests, the $z$ and $b$ parameters are assumed to be the same for all $J$ 
levels for a single velocity component, and the column density $N$ is the same for all transitions from a single $J$ level. Constraints delivered from the $J=0,1$ transitions are $\Delta\mu/\mu$ = (7.9$\pm$6.2$_{\rm stat}$)$\times10^{-6}$ and $\Delta\mu/\mu$ = (18.7$\pm$7.2$_{\rm stat}$)$\times10^{-6}$ for the 1\,VC model and 2\,VC model, respectively. If only the $J=2-4$ transitions are used, the 1\,VC and 2\,VC model deliver $\Delta\mu/\mu$ = (17.7$\pm$6.1$_{\rm stat}$)$\times10^{-6}$ and $\Delta\mu/\mu$ = (14.1$\pm$6.0$_{\rm stat}$)$\times10^{-6}$, respectively. As it can be expected in the case of underfitting, the 1\,VC constraint from the $J=0,1$ transitions seems to be slightly off from the other constraints, including those from the generic fits. 

\subsubsection{Separating Lyman and Werner transitions}
\label{lyman-werner-test}
A test is performed where a $\Delta\mu/\mu$ constraint is delivered from either only Lyman band or only Werner band transitions. For the 1\,VC model we find $\Delta\mu/\mu$ = (15.5$\pm$4.5$_{\rm stat}$)$\times10^{-6}$ from the Lyman transitions and $\Delta\mu/\mu$ = (14.3$\pm$17.7$_{\rm stat}$)$\times10^{-6}$ from the Werner transitions. For the 2\,VC model it is, respectively, $\Delta\mu/\mu$ = (15.9$\pm$4.6$_{\rm stat}$)$\times10^{-6}$ and $\Delta\mu/\mu$ = (10.1$\pm$17.7$_{\rm stat}$)$\times10^{-6}$. The derived constraints are in good agreement with each other and with the generic fit results. However the uncertainty of the constraints from Werner transitions is much larger than the one from Lyman transitions because there are around 4 times fewer Werner transitions in the fit. What the dominance of Lyman transitions over the fiducial result can imply is discussed in \ref{long-range-systematics}. Note again that the tests are performed not by excluding unwanted transitions but by leaving their $\Delta\mu/
\mu$ parameter fixed to zero.

\subsubsection{Excluding problematic regions}
Seventeen H$_2$ transitions are blended with narrow lines, some of which are due to known metal absorbers while others are unidentified. If the interloping lines are fitted inadequately, an effect in $\Delta\mu/\mu$ can be expected. To test this presumption, a fit without the regions containing these transitions is performed. For the 1\,VC and 2\,VC models the fitted subsample delivered, respectively, $\Delta\mu/\mu$ = (18.3$\pm$4.8$_{\rm stat}$)$\times10^{-6}$ and $\Delta\mu/\mu$ = (15.9$\pm$4.9$_{\rm stat}$)$\times10^{-6}$. The resulting constraints are in agreement with those from the generic fit which leads to a conclusion that contamination by metallic transitions does not affect positions of the H$_2$ transitions substantially. 
 
\subsubsection{Untying free parameters among different $J$ levels}
In the tests described above, redshifts and widths of the H$_2$ transitions were assumed to be the same, independent from $J$ level. To test the validity of this assumption, the low- and high-$J$ transitions are decoupled from each other, and allowed to assume different $z$ and $b$ values to verify if the two groups deliver consistent constraints. One possibility why it may not be the case is a spatial inhomogeneity in temperature in the absorbing cloud. Also, by performing such test we can test the possibility of underfitted velocity structure similarly as in \ref{low-high-J-test}. For the 1\,VC and 2\,VC models this test delivered, respectively, $\Delta\mu/\mu$ = (15.6$\pm$4.5$_{\rm stat}$)$\times10^{-6}$ and $\Delta\mu/\mu$ = (15.5$\pm$4.6$_{\rm stat}$)$\times10^{-6}$. These constraints agree within uncertainties with those from the generic fits which means that in the case of a relatively simple velocity structure of H$_2$, releasing some assumptions may not affect the $\Delta\mu/\mu$ considerably.

\subsubsection{Fitting parts of the spectrum}
\label{part-fitting}
In the presence of a long-range monotonic distortion of the wavelength scale, a significant effect on $\Delta\mu/\mu$ can potentially be generated, especially if only Lyman or only Werner transitions were fitted since the transitions in each band have monotonically increasing $K_{i}$ coefficients towards the blue wavelengths. The degeneracy is possibly broken if both Lyman and Werner transitions are fitted in a common wavelength interval because, e.g. $K_{i}$(Werner) = $-0.01$ while $K_{i}$(Lyman) = 0.03 at 1010 \textrm{\AA} (see Fig. \ref{fig1}). On the other hand, the range where only Lyman transitions are present has a better S/N ratio and, therefore, can have a substantial weight in the comprehensive fit.

We separate and fit the part of the spectrum where both Lyman and Werner transitions are present, i.e. bluewards from 3714 \textrm{\AA}, and as a complementary test we also fit the excluded Lyman transitions (redwards from 3714 \textrm{\AA}). For the 1\,VC test, the former test results in $\Delta\mu/\mu$ = (12.1$\pm$6.6$_{\rm stat}$)$\times10^{-6}$ and the latter $\Delta\mu/\mu$ = (15.4$\pm$9.9$_{\rm stat}$)$\times10^{-6}$. For the 2\,VC test, the results are $\Delta\mu/\mu$ = (11.1$\pm$6.7$_{\rm stat}$)$\times10^{-6}$ and $\Delta\mu/\mu$ = (17.9$\pm$10.0$_{\rm stat}$)$\times10^{-6}$, respectively. 
If the $\mu$ dependence of Werner transitions is removed, the fit (in the blue part) delivers $\Delta\mu/\mu$ = (10.5$\pm$6.7$_{\rm stat}$)$\times10^{-6}$ and $\Delta\mu/\mu$ = (9.7$\pm$6.8$_{\rm stat}$)$\times10^{-6}$ for the 1\,VC and 2\,VC models, respectively.

Generally, the constraints quoted above do not diverge significantly from those in the generic fit, however, there is a bias all of them can be affected by. In \ref{lyman-werner-test}, it was demonstrated that the constraints derived from generic model are dominated by the Lyman transitions because of their disproportionate contribution to the data set. In the test where we disregard Lyman transitions in the red part (38 in total), the Lyman-to-Werner ratio is reduced to 2:1 compared to 4:1 in the generic fit. Thus, this approach should be less biased than the generic one but it is still dominated by the Lyman transitions.

In Section \ref{long-range-systematics} the possibility of long-range distortions is explored further.

\section{Systematic uncertainty}

\subsection{Calibration residuals}
\label{sys-uncertainty-1}
One of the sources of systematic uncertainty is wavelength calibration residuals of ThAr lines. Typically, the residuals have a rms of 70 m s$^{-1}$, that is, at any given place in spectrum the wavelength scale is accurate within 70 m s$^{-1}$. Usually, more than 10 ThAr lines are detected and multiple H$_2$ transitions are fitted in each echelle order; thereby the effect of any systematic trends is reduced to 30 m s$^{-1}$ at most \citep{murphy2007}. Given a span in $K_i$ of 0.05, a velocity shift of 30 m s$^{-1}$ translates into a $\Delta\mu/\mu$ shift of 2.0$\times$10$^{-6}$, applying the following relation: $\Delta v = c \Delta K_{i}\Delta\mu/\mu$.

\subsection{Non-attached ThAr}
The UVES spectrograph is designed in such way that the grating is repositioned between different exposures. Although the repositioning of the grating should be accurate to within 0.1 pixel \citep{dodorico2000}, the ThAr calibrations can be obtained immediately after science observation, without initiating grating reset, so that potential uncertainties in the wavelength scale are avoided. The bulk of our exposures have attached ThAr calibrations (60 $\%$ of the data). To test for possible miscalibration effects, we fit a `subspectrum' -- a spectrum comprising of only a subset of exposures but composed by the same process as the full spectrum -- which includes only these `well-calibrated' exposures. For the S/N ratio reduced from 35 to 32 at 370 nm, the 1\,VC model delivers $\Delta\mu/\mu$ = (16.7$\pm$4.9$_{\rm stat}$)$\times10^{-6}$ while the 2\,VC model  $\Delta\mu/\mu$ = (16.4$\pm$4.9$_{\rm stat}$)$\times10^{-6}$ (also see Fig. \ref{fig4}). It can be concluded that the effect on $\Delta\mu/\mu$ due to data 
with non-attached ThAr is not evident in our case. However, this conclusion cannot be generalized and applied to every UVES spectrum, e.g. in the case of varying $\alpha$ studies where just a few transitions are fitted in the spectrum, the effect -- if present -- might cause considerable shifts.

\subsection{Intra-order wavelength scale distortions}
\label{sys-uncertainty-3}
\cite{whitmore2010} showed that intra-order distortions (i.e. distortions of the wavelength scale within echelle orders which repeat from order to order) up to $\sim$100 m s$^{-1}$ can be expected in VLT/UVES spectra. Thus, a $+/-$100 m s$^{-1}$ saw-tooth wavelength distortion is introduced to each echelle order of the B0642$-$5038 spectrum using \textsc{uves$\_$popler} before they are combined into a single spectrum for analysis in \textsc{vpfit}. Then, taking all 111 transitions into account, we perform a fit on a spectrum with the introduced intra-order wavelength scale distortions. Results of such tests are model-dependent, but generally intra-order distortions are not expected to be a significant problem for $\mu$ variation analysis, because the H$_2$ transitions are spread over multiple orders. From the test which we performed on a distorted spectrum, we find a contribution to the systematic uncertainty of $\Delta\mu/\mu$ at the level of 0.6$\times$10$^{-6}$ (see Fig. \ref{fig4}). The effect we find 
here 
is of the same order as in previous studies where such an analysis was also performed \citep{malec2010a, king2011, weerdenburg2012}.

\subsection{Uncertainty from spectral re-dispersion}
\label{sys-uncertainty-4}
To compose a final 1-D spectrum, several echelle orders are combined by re-dispersing them onto a common wavelength scale and taking their weighted mean. The re-binning may cause flux correlations between neighbouring pixels. Thus, the choice of a wavelength grid can in principle affect the measurement of $\Delta\mu/\mu$. To test this, we choose several slightly different grids in the range from 2.3 to 2.7 km s$^{-1}$ per pixel. The maximal deviation among the resulting $\Delta\mu/\mu$ values is $\pm3.0\times10^{-6}$. This is added to the total systematic uncertainty budget. 
Previous studies of different H$_2$ absorbers report deviations in $\Delta\mu/\mu$ due to re-dispersion in the range from 0.2$\times10^{-6}$ \citep{weerdenburg2012} to 0.8$\times10^{-6}$ \citep{malec2010a} for pixel sizes $\sim$1.3\,km\,s$^{-1}$ and 1.4$\times10^{-6}$ \citep{king2011} for $\sim$2.5\,km\,s$^{-1}$ pixels like those in the present analysis.

\subsection{Long-range wavelength scale distortions}
\label{long-range-systematics}
The possibility of long-range distortions between the ThAr calibration spectra and quasar spectra recorded with VLT/UVES has been reported recently by \citep{rahmani2013}. In a study focusing on a $\mu$-variation analysis from H$_2$ absorption towards HE0027--1836, \cite{rahmani2013} report significant long-range calibration errors for UVES data, especially strong in exposures taken in 2012. Wavelength distortion errors, when using ThAr spectra for calibration could, in principle, be produced by differing beam paths and/or slit illumination distributions between the quasar science exposure and the subsequent ThAr calibration exposure.
After reflecting from the primary, secondary and tertiary mirrors, quasar light is directed through a derotator and into the UVES enclosure.
Once it enters the enclosure, the beam is split by a dichroic into the two arms of the spectrograph: a blue arm and a red arm. 
Each beam passes through its respective optics: a blue (red) slit, the cross dispersers, several mirrors, and finally lands on the blue (red) CCD(s).
The light from the calibration lamp is sent into the UVES enclosure by reflecting from a calibration mirror that is slid into the optical path of the telescope. 
After entering the enclosure, it interacts with the same optics described above.
However, there could be an angular offset between the science and calibration exposures, which would result in slightly differing beam paths through the spectrometer, leading to a possible wavelength distortion. Also, each slit is illuminated fully by the ThAr lamp, while the quasars are unresolved point sources. Therefore, the quasar and ThAr light may produce different point-spread functions on the CCDs and, if that difference varies across the CCDs, a long-range distortion between the quasar and ThAr calibration may result.

\cite{rahmani2013} used a cross-correlation technique to compare UVES spectra of various asteroids taken across several years with a solar spectrum recorded with a Fourier Transform Spectrometer (FTS). They found long-range velocity slopes which they translated into systematic offsets for $\Delta\mu/\mu$ lying in the range between $2.5 \times 10^{-6}$ and $13.3 \times 10^{-6}$. Prompted by the conclusions of \cite{rahmani2013} we will later assess long-range wavelength distortions in the B0642$-$5038 spectrum using the asteroid method.

We first attempt to address the likely sign and magnitude of long-range distortions in the actual B0642$-$5038 spectra using simulations. In Section \ref{part-fitting} we discussed that the Werner and Lyman transitions, when fitted simultaneously, can break a possible degeneracy between long-range distortions and $\Delta\mu/\mu$. To illustrate this further we use simulated spectra with distortions introduced following the formalism presented by \cite{malec2010a}, Section 4.2.2 in their paper. The wavelength scale is compressed to simulate a long-range distortion and the H$_2$ transitions are shifted to mimic a non-zero ($\Delta\mu/\mu)_{\rm sys}$. We used the 1\,VC generic model to produce simulated spectra with the same S/N ratio as the real quasar spectrum, three different noise realisations in total. Each of these three simulated spectra was distorted to mimic ($\Delta\mu/\mu)_{\rm sys}$ of $-17$ and $+17 \times 10^{-6}$, and fitted using the 1\,VC generic model including all 111 H$_2$ transitions. As 
expected, the $\Delta\mu/\mu$ values returned from fitting agree with ($\Delta\mu/\mu)_{\rm sys}$ within uncertainties, also when the undistorted simulations (i.e. ($\Delta\mu/\mu)_{\rm sys}$ = 0) were analyzed. In other words, in a relationship $\Delta\mu/\mu = s_1\times$($\Delta\mu/\mu)_{\rm sys}$+$b_1$ the coefficient $s_1$ is close to unity, and $b_1$ is of the order of $\Delta\mu/\mu$ statistical uncertainty.

In a second step, the same simulated spectra with long-range distortions were used to derive $\Delta\mu/\mu$ constraints separately from the blue and red parts (with a dividing line at 3714\,\textrm{\AA} as in Section \ref{part-fitting}). To compare the constraints returned from the blue part with those from the red part we use a relationship $(\Delta\mu/\mu)_{\rm red} - (\Delta\mu/\mu)_{\rm blue}  = s_2\times$($\Delta\mu/\mu)_{\rm sys}$+$b_2$. In this case, a $s_2$ coefficient close to unity means that the blue part is very resistant against the long-range distortions represented by ($\Delta\mu/\mu)_{\rm sys}$. From our simulated spectra we derive $s_2\simeq0.8$. The real quasar spectrum fitted with a 1\,VC model returns $(\Delta\mu/\mu)_{\rm red} - (\Delta\mu/\mu)_{\rm blue} = (3.3\pm11.9)\times10^{-6}$ (see Section \ref{part-fitting}). Assuming $b_2$ is negligible (few $\times10^{-6}$), we can estimate the size of a possible linear wavelength distortion in the real spectrum: ($\Delta\mu/\mu)_{\rm sys} =  (
3.3\pm11.9)\times10^{-6}/0.8 = (4 \pm 15)\times10^{-6}$. The possible distortion itself is not large but it has a large uncertainty which makes the interpretation difficult. 

A different way of quantifying long-range distortions is to observe objects with a well-understood spectrum which can then be compared with a reference spectrum.
An iodine cell has been used to quantify the calibration differences between the ThAr wavelength solution of quasar spectra observed with the iodine cell in the line of sight and a reference FTS iodine cell spectrum \citep{griest2010, whitmore2010}. 
We use a new implementation of this `supercalibration` method (Whitmore et al., in prep.), which is similar to the iodine cell. The only practical difference is instead of using the iodine cell spectrum as a reference, we use the solar spectrum.  
The general procedure for finding a velocity shift $v_{\rm shift}$ between the ThAr solution and the FTS reference spectrum begins by using a science exposure that has been calibrated in the standard way that our a quasar exposures were calibrated: 1) take a science exposure, 2) take a ThAr calibration exposure, 3) solve for the ThAr wavelength solution.
After these steps are completed, the supercalibration technique solves for relative velocity shifts between the science exposure's ThAr wavelength solution and a reference spectrum. 

The details of how we implement this supercalibration technique for this paper are as follows. We use the FTS solar spectrum KPNO2010 detailed in \cite{chance2010}, and is publicly available\footnote{\url{http://kurucz.harvard.edu/sun/irradiance2005/irradthu.dat}} as our reference spectrum. We chose an archival exposure of the asteroid Ceres taken in December 2007, a few days before the beginning of the quasar observations (which span the next two months, see Table \ref{observations}). Also we used a number of the solar twin HD28099 and HD76151 spectra observed a month later than Ceres, conveniently close in time to a bulk of the quasar observations (see Fig. \ref{fig:time-dist}). A 500 km s$^{-1}$ segment of the spectrum is taken from the asteroid/solar twin science exposure, and an overlapping region of the solar FTS spectrum is taken as the model. 
The model is modified by the following 5 parameter transformation: 1) a single velocity shift, 2) a multiplicative flux scaling factor, 3) an additive flux offset factor, 4) the sigma width of a symmetric Gaussian instrument profile, and 5) a linear continuum slope correction.
We minimize the $\chi^2$ between the model and the data.
The wavelength ThAr corrections found with this supercalibration technique are reported as $v_{\rm shift}$ with the following sign conventions:
\begin{equation}
    \lambda_{\rm shift} = \lambda_{\rm reference} - \lambda_{\rm ThAr},
\end{equation}    

\begin{equation}
    v_{\rm shift} = c \times \frac{\lambda_{\rm shift}}{\lambda}.
\label{eq:vshift}
\end{equation}

\begin{center}
  \begin{figure}
    \includegraphics[width=0.95\textwidth]{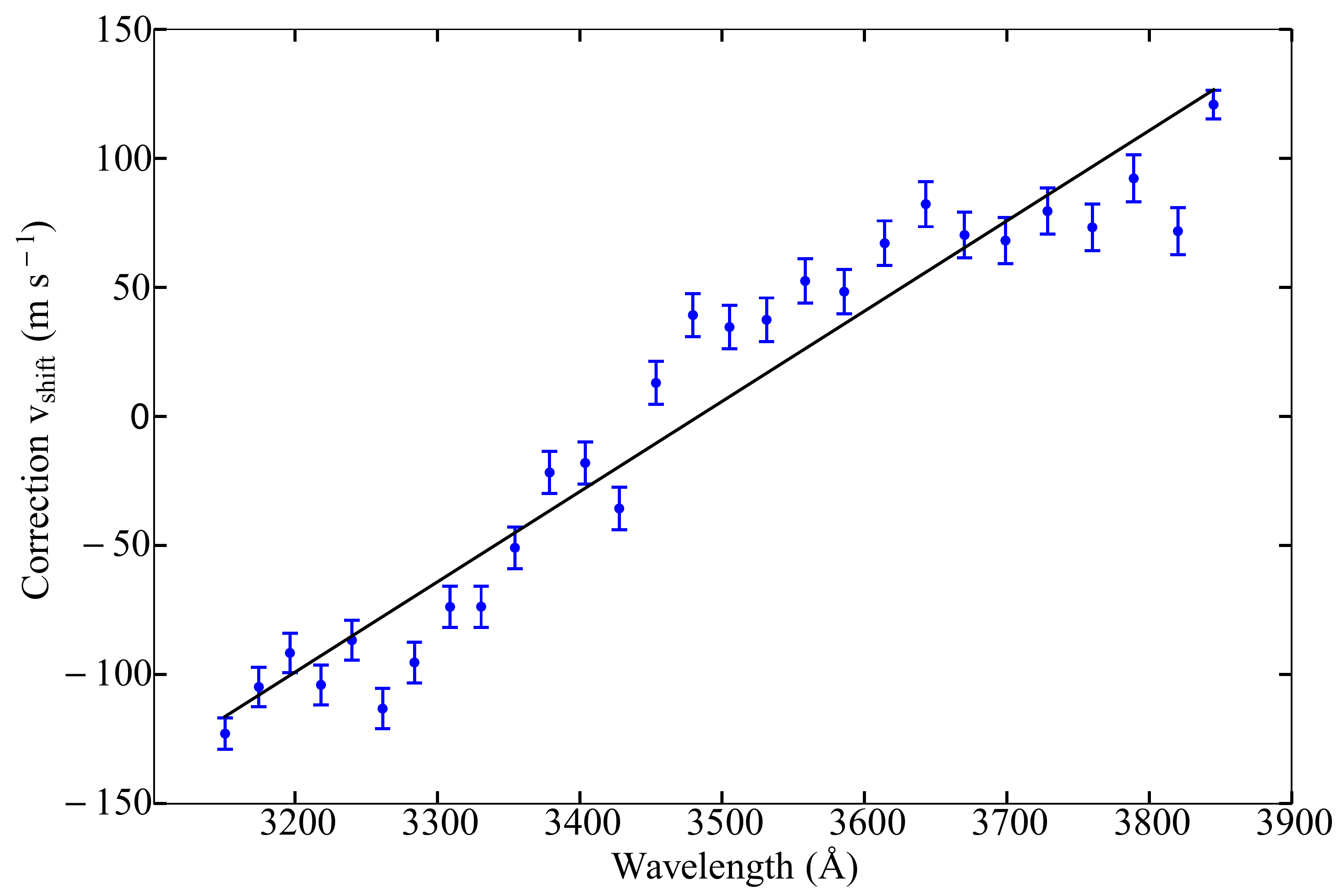}
    \caption{A plot of the long-range wavelength distortion found by the supercalibration technique in the Ceres 2007 exposure (orders 5--33). The $v_{\rm shift}$, as defined in Eq.~\ref{eq:vshift}, is the correction that needs to be applied to the ThAr wavelength solution to align with the fiducial FTS solar spectrum.
    We plot the average $v_{\rm shift}$ for each echelle order, and the size of the error bar is the standard deviation of the $v_{\rm shift}$ within that order. 
    The plot has been shifted by a constant velocity to account for radial velocity and slit-offset effects. In other words, the absolute scale is not meaningful, but the relative scale is informative. The equation of the best fit line is $v_{\rm shift}(\lambda) = {\rm A} \times \lambda + {\rm B}$, with A = 350 m s$^{-1}$ per 1000 \AA, and B = $-1218.0$ m s$^{-1}$.
    \label{fig:std-slope}}
  \end{figure}  
\end{center}

\begin{center}
  \begin{figure}
    \includegraphics[width=0.95\textwidth]{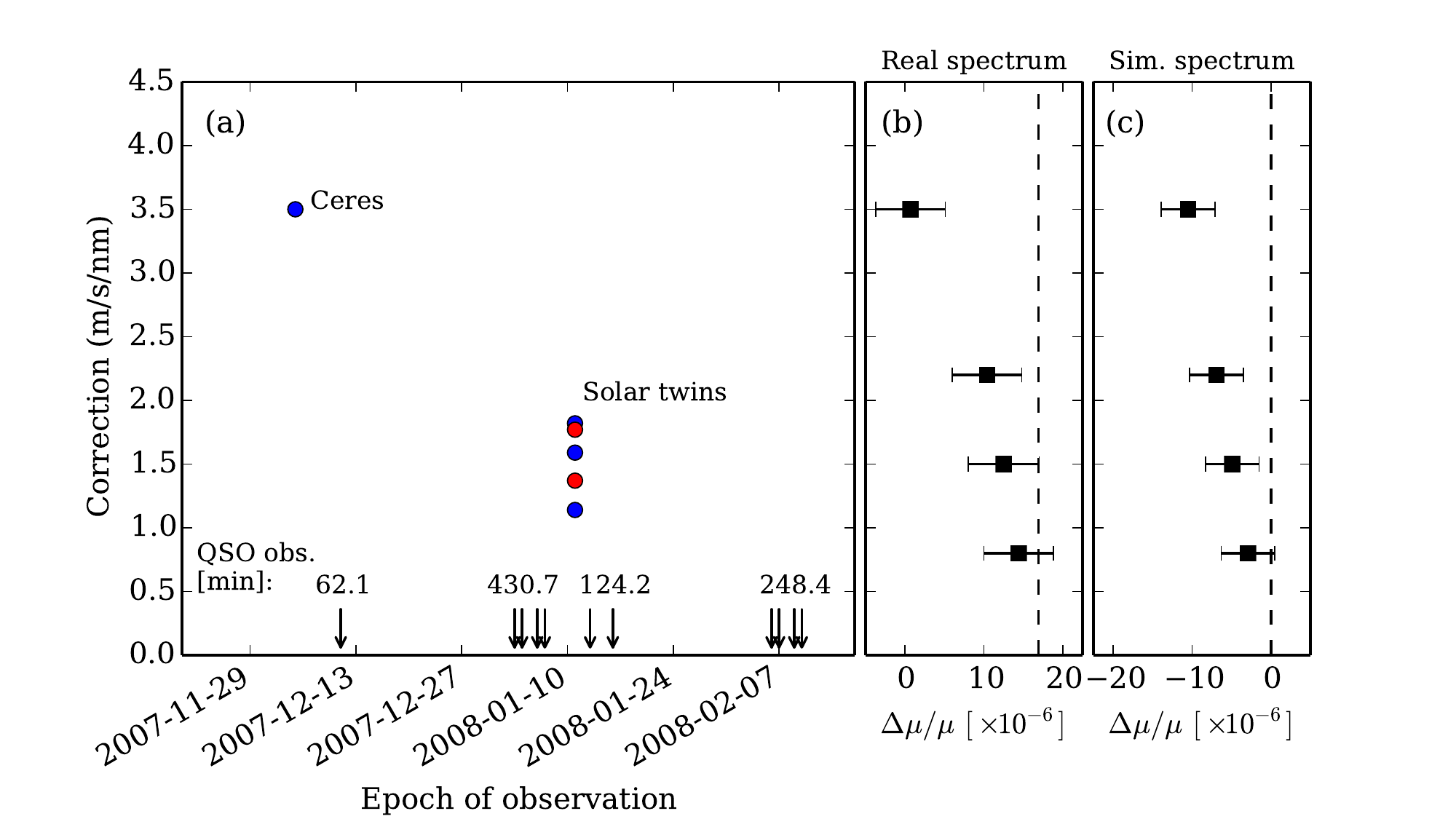}
    \caption{(a) Corrections of the wavelength scale, as derived from observations of Ceres and solar twins, are displayed versus the observing epoch. Blue (red) points refer to corrections of the wavelengths at $<$450 nm ($>$450 nm). Quasar observations from program 080.A-0288(A) are marked on the time axis; most of them were conducted around the time when the solar twin spectra were taken. (b) The data points refer to the $\Delta\mu/\mu$ constraints as derived from the B0642$-$5038 spectrum when wavelength corrections are applied to it (1 VC model). The vertical dashed line shows a $\Delta\mu/\mu$ constraint from an uncorrected spectrum. Applying wavelength corrections results in $\Delta\mu/\mu$ values which deviate from zero by less than the one from the uncorrected spectrum. Note that we apply wavelength corrections to the total averaged spectrum and not to the individual exposures. (c) A simulated quasar spectrum is skewed using the same wavelength corrections as in (b). Corrections of this sign result 
in 
a $\Delta\mu/\mu$ measurement being smaller than the input $\Delta\mu/\mu = 0$ (indicated by the dashed vertical line). A correction of 3.5 m\,s$^{-1}$\,nm$^{-1}$ (implied by Ceres observations) leads to a $\Delta\mu/\mu$ shift from 0 to $-10.5\times10^{-6}$. A correction based on the solar twin (1.5 m\,s$^{-1}$\,nm$^{-1}$) results in a $\Delta\mu/\mu$ shift from 0 to $-4.9\times10^{-6}$. The uncertainties of the $\Delta\mu/\mu$ constraints are defined by the S/N of the simulated spectrum.
    \label{fig:time-dist}}
  \end{figure}  
\end{center}

The supercalibration technique is sensitive to both short and long range distortions. Since the contribution to the systematic error budget of short range (intra-order) distortions are quantified in \ref{sys-uncertainty-3}, the long-range distortions remain to be incorporated. A plot of the average $v_{\rm shift}$ per order is plotted, with an error bar that is the size of the standard deviations is shown in Fig.~\ref{fig:std-slope}. We fit a linear model of the long-range distortion and use this model in our analysis.

As it can be seen from Fig. \ref{fig:time-dist}(a), a correction derived from the Ceres observations is twice as big as the one from the solar twins. We apply these two corrections to the spectrum of B0642$-$5038 (1 VC model) and derive, respectively, $\Delta\mu/\mu$ of ($0.7\pm4.4_{\rm stat}$)$\times10^{-6}$ and ($12.5\pm4.4_{\rm stat}$)$\times10^{-6}$. These values, especially the former, differ substantially from an uncorrected measurement ($\Delta\mu/\mu$ = (16.9$\pm$4.4$_{\rm stat}$)$\times10^{-6}$). As an additional test, we apply the corrections to a simulated spectrum. Similarly as before, we used the 1\,VC generic model to produce a simulated spectrum with the same S/N ratio as the real quasar spectrum, and imposed $\Delta\mu/\mu = 0$. Fitting this spectrum with two different correction values results in $\Delta\mu/\mu$ of ($-10.5\pm3.4_{\rm stat}$)$\times10^{-6}$ (larger correction from Ceres) and ($-4.9\pm3.4_{\rm stat}$)$\times10^{-6}$ (smaller correction from the solar twins). In both cases -- 
when the corrections are applied to the real and simulated spectrum -- a measured $\Delta\mu/\mu$ value is smaller than from a corresponding uncorrected spectrum (see Fig. \ref{fig:time-dist} (b) and (c)). In other words, if the long range wavelength distortions are neglected, a measured $\Delta\mu/\mu$ will likely be more positive than its actual value. A number of asteroid calibrations taken with UVES over surrounding years agree on the sign of the effect. These offsets are relatively large, exceeding the estimates of statistical uncertainties and other systematic effects \citep{rahmani2013}. However, an important question to be addressed when making a correction of $\Delta\mu/\mu$ is that of accuracy of the correction. As can be seen from Fig. \ref{fig:time-dist}(a), most of the quasar exposures are taken within $\pm$8 days from the solar twins' observations, and thus we further rely on the solar twins in making an adjustment of the fiducial $\Delta\mu/\mu$ measurement, i.e. 
we use $\Delta\mu/\mu = -4.4\times10^{-6}$ as a correction. 
Note also that the same central wavelengths (setting of the dichroic) have been adopted in the observations of the solar twins and quasar while that is not the case for the Ceres asteroid (see Table \ref{observations}). Nevertheless, we include the latter measurement in the estimation of the $\Delta\mu/\mu$ correction accuracy which we base on a spread in the derived wavelength correction values over a month's time (Fig. \ref{fig:time-dist}(a)). A standard deviation of the averaged correction values equals $\pm$0.7 m\,s$^{-1}$\,nm$^{-1}$ and thus we fit spectra corrected by $\pm$0.7 m\,s$^{-1}$\,nm$^{-1}$ with respect to the correction from the solar twins (1.5 m\,s$^{-1}$\,nm$^{-1}$). These corrections result in $\Delta\mu/\mu$ shifts of $\pm$2.0$\times10^{-6}$ with respect to the results derived from both the real quasar spectrum and the simulated spectrum. 
Thus, we further add $\pm$2.0$\times10^{-6}$ to the systematic error budget of $\Delta\mu/\mu$.

\subsection{Summary of systematic uncertainty}
From the performed tests we estimate four definite contributions to the total systematic error of $\Delta\mu/\mu$: the calibration residuals can contribute up to 2.0$\times$10$^{-6}$, the intra-order distortions can introduce an error of 0.6$\times$10$^{-6}$, an effect of 0.7$\times$10$^{-6}$ can be expected due to non-attached ThAr calibrations, while the spectral redispersion may introduce an error of 3.0$\times$10$^{-6}$. 
Adding these four contributions in quadrature, we obtain the total systematic error of $\Delta\mu/\mu = 3.7\times$10$^{-6}$. The resulting constraint is then $\Delta\mu/\mu = (17.1 \pm 4.5_{\rm stat} \pm3.7_{\rm sys})\times10^{-6}$. Analysis of solar twin and asteroid spectra show that an additional systematic effect pertaining to wavelength miscalibration over long ranges is found in the quasar spectrum. Due to this miscalibration the $\Delta\mu/\mu$ measurement may have been shifted towards more positive values. While both the asteroid and solar twin calibrations suggest corrections of the same sign, their magnitudes differ by a factor of 2. In making a correction, we rely on the solar twin spectra which are taken closer in time to most of the quasar observations and with the same spectrograph settings. It delivers a $\Delta\mu/\mu$ correction of $(-4.4\pm2.0_{\rm sys})\times$10$^{-6}$, and hence a corrected $\Delta\mu/\mu$ of $(12.7 \pm 4.5_{\rm stat} \pm4.2_{\rm sys})\times10^{-6}$.

\begin{figure}
\centering
 \includegraphics{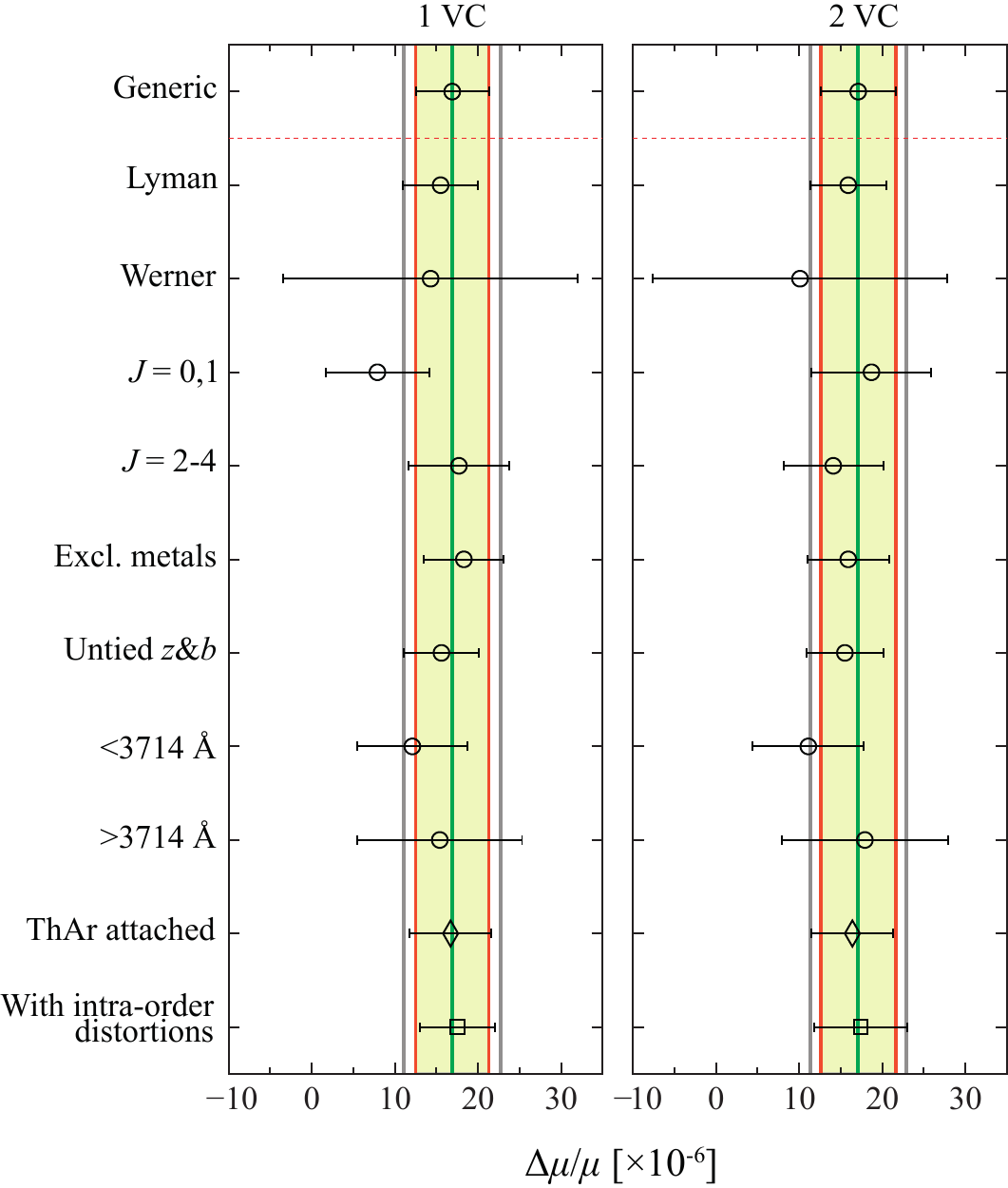}
\caption{Constraints from various tests described in Section \ref{tests} are displayed for comparison with the one from a generic model which includes all 111 transitions and all available spectral data. The vertical green line in each panel indicates the $\Delta\mu/\mu$ value from the generic test. The shaded area and red (gray) vertical lines indicate the 1$\sigma$ statistical (statistical and systematic) uncertainties associated with the constraints from the generic tests. The two bottom constraints are used to estimate systematic error on $\Delta\mu/\mu$ due to potential wavelength scale inaccuracies described in Sections \ref{sys-uncertainty-1}-\ref{sys-uncertainty-4}.}
\label{fig4}
\end{figure}

\section{Discussion}
In the present study ESO-archival spectra toward a quasar system B0642$-$5038 with a DLA at redshift 2.659 are analyzed to extract information on a possible variation of the proton-electron mass ratio $\mu$. The DLA contains a molecular fraction of $\log f= -2.18 \pm 0.08$ in which 111  H$_2$ lines were identified usable for a $\mu$-variation analysis. The spectrum is of good quality and has a signal-to-noise ratio of $\sim 35$ in the relevant wavelength region of H$_2$ absorbers, which is just a bit lower than some other systems that were analyzed previously: J2123$-$0050 \citep{malec2010a, weerdenburg2012}, Q0528$-$250 \citep{king2008, king2011}, Q0405$-$443 \citep{king2008}, Q0347$-$383 \citep{king2008, wendt2012}, and better than Q2348$-$011 \citep{bagdonaite2012} and HE0027$-$1836 \citep{rahmani2013}.

The resulting value of $\Delta\mu/\mu = (17.1 \pm 4.5_{\rm stat} \pm 3.7_{\rm sys}) \times 10^{-6}$ represents a result which is in itself a remarkable 3$\sigma$ effect on a varying constant. However, we find evidence for a long-range distortion of the wavelength scale in the analyzed spectrum. Based on our analysis of asteroid and solar twin spectra and on a similar study by \cite{rahmani2013} we conclude that due to long-range distortions $\Delta\mu/\mu$ in this spectrum has shifted away from its actual value towards more positive values. While the sign of this systematic error seems to be consistent among data taken in different epochs, the amplitude appears to be varying. We use solar twin spectra which are taken close in time to the quasar observations to make a correction which reduces the significance of the initial measurement and delivers $\Delta\mu/\mu = (12.7 \pm 4.5_{\rm stat} \pm4.2_{\rm sys})\times10^{-6}$ or $(12.7 \pm 6.2)\times10^{-6}$ if the uncertainties are added in quadrature.

Regarding the analysis presented here, of great importance is the consistency found in the statistical and systematic tests performed on the data set. These tests, results of which are graphically displayed in Fig. \ref{fig4}, constitute velocity-component analysis, separate effect of subsets of Lyman and Werner bands, separate sets of populated rotational states associated with cold and warm molecular fractions, separation of wavelength regions, and effects of short-range (i.e. intra-order) distortions of the wavelength scale. No significant difference was found between the sets of quasar exposures with `attached' ThAr calibration spectra and the sets for which the ThAr calibration was performed at the end of the night after resets of the grating in the spectrometer.

\begin{figure}[!h]
 \includegraphics[scale=0.5]{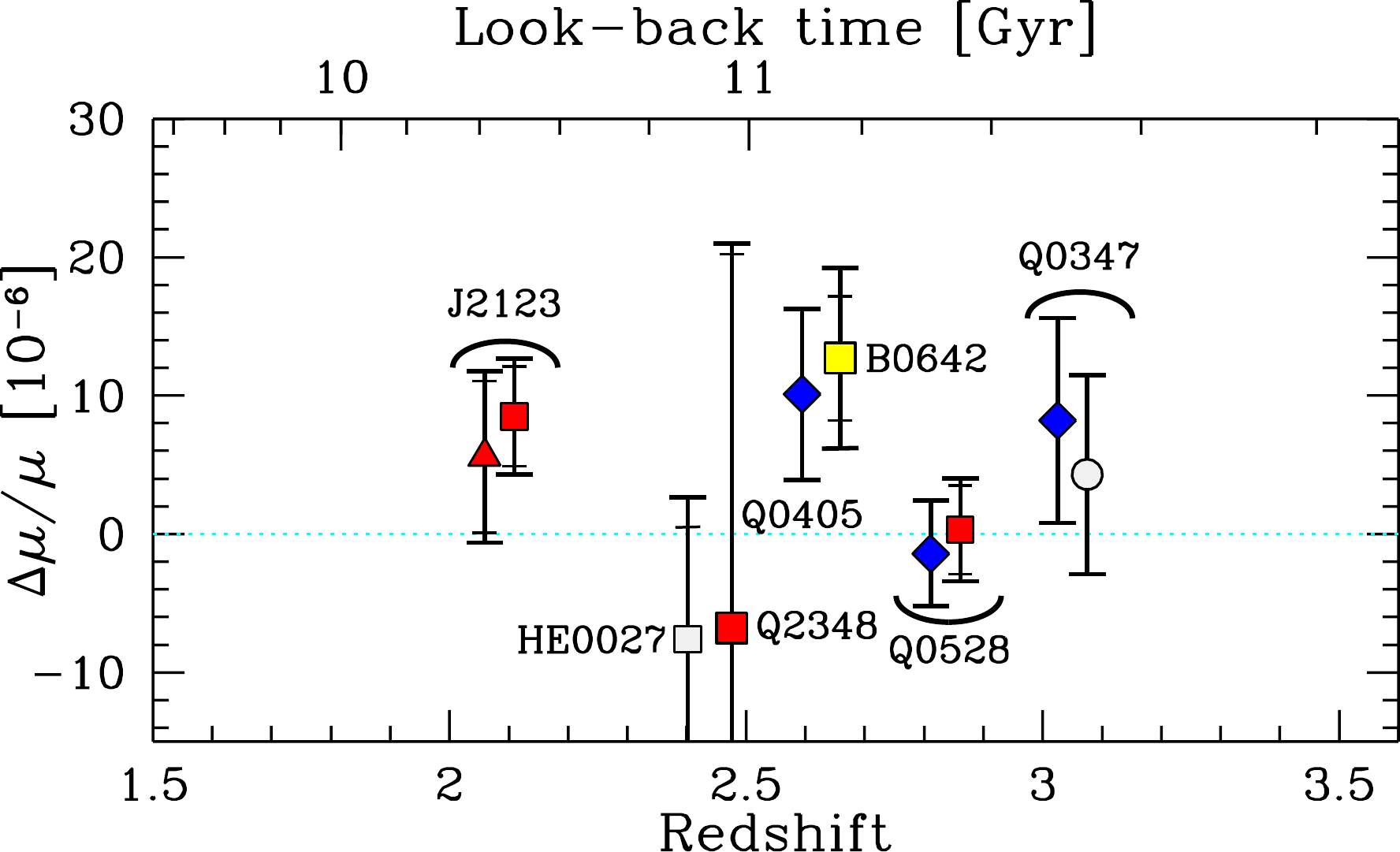}
 \caption{$\Delta\mu/\mu$ constraints from H$_2$ in 7 different quasar sightlines: Q0528$-$250 \citep{king2008, king2011}, J2123$-$0050 \citep{malec2010a, weerdenburg2012}, Q0347$-$383 \citep{king2008, wendt2012}, Q2348$-$011 \citep{bagdonaite2012}, Q0405$-$443 \citep{king2008}, HE0027$-$1836 \citep{rahmani2013}, and B0642$-$5038 (this work). Note that some absorbers where analyzed more than once -- in these cases one of the point is offset on $z$-scale by +0.05 to avoid overlap. Also the Q2348 point is offset by +0.05 to avoid overlap with the HE0027 which is located at a similar redshift. All the constraints shown here are derived by employing the comprehensive fitting method, except for a constraint from Q0347$-$383, shown as a grey circle \citep{wendt2012}.
}
\label{all-mu-constraints}
\end{figure}

The presented $\Delta\mu/\mu$ constraint can be viewed in the perspective of the entire set of results that is currently being produced for H$_2$ absorptions at high redshift. In Fig. \ref{all-mu-constraints} this dataset is plotted. From all the analyses being performed we have collected those results that were obtained through the comprehensive fitting method. In previous reports an extensive discussion is provided on the advantages of the comprehensive fitting over the line-by-line method \citep{malec2010a, king2008}. Only for the case where a single velocity component is definitive, such as in Q0347$-$383 \citep{wendt2012}, a result from a line-be-line analysis is included in the overview. Averaging over the presented dataset results in $\Delta\mu/\mu$=(4.0$\pm$1.8)$\times10^{-6}$ for look-back times in the range of 10-12 billion years.

Naturally, a question arises as to what extent the $\Delta\mu/\mu$ constraints measured in the past are affected by the long-range wavelength miscalibration. In the dataset displayed in Fig. \ref{all-mu-constraints}, only the constraints from the B0642$-$5038 and HE0027$-$1836 quasar sightlines are derived with the long-range wavelength distortion effect taken into account. According to the present study, wavelength distortions can vary in amplitude within one night as well as over a month's time, while \cite{rahmani2013} found year-to-year changes. Thus, regarding the H$_2$ absorbers which have been analyzed so far, a case-by-case re-analysis might be necessary, where this newly found systematic effect is investigated. Although from the presently available information it seems likely that the individual and average $\Delta\mu/\mu$ measurements are biased towards more positive values, the two examples discussed further reinforce the motivation for individual re-analysis using absolute calibration methods 
such 
as the supercalibration method. As for VLT/UVES, a detailed and very accurate study was made for the Q0528$-$250 H$_2$ absorbing system. Based on UVES data recorded in Jan.~2003, reported by~\citet{ivanchik2005}, \citet{king2008} performed a reanalysis with the comprehensive fitting method and retrieved a constraint of $\Delta\mu/\mu =(-1.4 \pm 3.9) \times 10^{-6}$. Based on an independent data set for Q0528$-$250, recorded in the period Nov.~2008 to Feb.~2009, \citet{king2008} deduced a constraint of $\Delta\mu/\mu =(0.3 \pm 3.7) \times 10^{-6}$, in perfect agreement, while the data sets of the same object, analyzed by the same methods were obtained over a time interval of 6 years. As it is not understood how the long-range wavelength distortions change with time, it remains to be seen whether it is just by coincidence that these two constraints were affected similarly. Further, there is a result on the analysis of H$_2$ absorption toward J2123$-$0050, from two different telescopes. While \cite{malec2010a} 
obtained $\Delta\mu/\mu =(5.6 \pm 5.6_{\rm stat} \pm 2.9_{\rm sys}) \times 10^{-6}$ from an observation of J2123$-$0050 with HIRES/Keck, \citet{weerdenburg2012} obtained $\Delta\mu/\mu =(8.5 \pm 3.6_{\rm stat} \pm 2.2_{\rm sys}) \times 10^{-6}$ from VLT/UVES, also in good agreement. Here it is noted that the same comprehensive fitting method was used in both analyses, and that, so far J2123$-$0050 is the best H$_2$ absorbing system analyzed, in terms of brightness of the background quasar and the column density of the H$_2$ absorbing galaxy. This particular absorber was investigated in a recent study by \cite{evans2013}, in which direct comparison of the HIRES/Keck and VLT/UVES spectra was made to detect possible relative velocity shifts between the two. While some indications of a constant offset between the spectra were found, no significant wavelength-dependent shift could be detected. The sensitivity of the direct comparison method is similar to that of the supercalibration approach used in the current 
study (a few m\,s$^{-1}$\,nm$^{-1}$). Thus, the results by \cite{evans2013} might be held as counterevidence against long-range distortions of the UVES wavelength calibration, unless HIRES/Keck suffers from a similar problem. More investigations of this phenomenon are urgently needed, in particular for studies as the present one that require understanding of the calibration of the spectrometer at its extreme limits. This would lead to establishing a firm constraint on a variation of $\mu$ at redshifts $z \sim 2-3$.

Studies at lower redshifts, based on the ammonia and methanol methods, yield a constraint of $\mu$ varying at the level of less than $3 \times 10^{-7}$ \citep{henkel2009, kanekar2011, bagdonaite2013a, bagdonaite2013b}. These strongly constraining findings, produced from absorbing clouds toward PKS1830$-$211 and B0218$+$357, may be interpreted as contradictive to less constraining results beyond $z > 2$. In view of physical models linking coupling strengths to the ratio of matter vs. dark energy in the universe, where variation of constants is frozen by dark energy \citep{sandvik2002, barrow2002}, it remains of importance to search for drifting constants in various evolutionary stages of the Universe, i.e. at different redshifts.

\clearpage
\newpage




\acknowledgments



{\it Facilities:} \facility{VLT (UVES)}

\clearpage



\end{document}